\documentclass[useAMS,usenatbib,aastex]{mnras}

\usepackage{graphicx}
\usepackage{amssymb,amsmath,pdflscape}
\usepackage{natbib}
\usepackage{auto-pst-pdf}

\newcommand{\acal}{\mathbfit{f}}
\newcommand{\bcal}{\mathbfit{u}}

\newcommand{\elik}{{\mathbf K}}
\newcommand{\elie}{{\mathbf E}}
\newcommand{\elil}{{\mathbf L}}

\newcommand{\Psigamma}{\Psi_{\cal C}}
\newcommand{\psif}{\Psi_p}
\newcommand{\gf}{ - \left. \partial_R \Psi \right|_{\rf}}
\newcommand{\kc}{{k_c}}
\newcommand{\kpc}{{k'_c}}
\newcommand{\etac}{\eta_c}
\newcommand{\rc}{R_c}
\newcommand{\zc}{Z_c}
\newcommand{\rf}{{R_{\rm p}}}

\newcommand{\bin}{{b_{\rm in}}}
\newcommand{\bout}{{b_{\rm out}}}
\newcommand{\pin}{{p_{\rm in}}}
\newcommand{\pout}{{p_{\rm out}}}

\begin{document}

\title{Interior potential of a toroidal shell from pole values}
\author[J.-M. Hur\'e et al.]
{J.-M. Hur\'e$^{1,2}$\thanks{E-mail:jean-marc.hure@u-bordeaux.fr},
  A. Trova$^{3}$,
  V. Karas$^{4}$,
and C. Lesca$^{1}$\\
$^{1}$Univ. Bordeaux, LAB, UMR 5804, F-33615, Pessac, France\\
$^{2}$CNRS, LAB, UMR 5804, F-33615, Pessac, France\\
$^{3}$University of Bremen, Center of Applied Space Technology and Microgravity (ZARM), 28359 Bremen, Germany\\
$^{4}$Astronomical Institute, Academy of Sciences, Bo\v{c}n\'{\i} II 1401, CZ-14100 Prague, Czech Republic}

\date{Accepted 2019 April 24. Received 2019 April 6; in original form 2019 January 16}
 
\pagerange{\pageref{firstpage}--\pageref{lastpage}} \pubyear{???}

\maketitle

\label{firstpage}

\begin{abstract}
We have investigated the toroidal analog of ellipsoidal shells of matter, which are of great significance in Astrophysics. The exact formula for the gravitational potential $\Psi(R,Z)$ of a shell with a circular section at the pole of toroidal coordinates is first established. It depends on the mass of the shell, its main radius and axis-ratio $e$ (i.e. core-to-main radius ratio), and involves the product of the complete elliptic integrals of the first and second kinds. Next, we show that successive partial derivatives $\partial^{n +m} \Psi/\partial_{R^n} \partial_{Z^m}$ are also accessible by analytical means at that singular point, thereby enabling the expansion of the interior potential as a bivariate series. Then, we have generated approximations at orders $0$, $1$, $2$ and $3$, corresponding to increasing accuracy. Numerical experiments confirm the great reliability of the approach, in particular for small-to-moderate axis ratios ($e^2 \lesssim 0.1$ typically). In contrast with the ellipsoidal case (Newton's theorem), the potential is not uniform inside the shell cavity as a consequence of the curvature. We explain how to construct the interior potential of toroidal shells with a thick edge (i.e. tubes), and how a core stratification can be accounted for. This is a new step towards the full description of the gravitating potential and forces of tori and rings. Applications also concern electrically-charged systems, and thus go beyond the context of gravitation.
\end{abstract}


\section{Introduction}

As elementary constituents of any model of slowly rotating star, thin homoeoids, i.e. infinitely thin ellipsoidal shells, are of major mathematical importance in Astrophysics \citep[e.g.][]{binneytremaine87}. They are equipotential surfaces, as revealed already by Newton's findings \citep{kellogg29}. In addition, the net force inside any solid ellipsoid (made of self-similar homoeoids), perpendicular to its main axis, happens to be linear with the cylindrical radius $R$, exactly as the centrifugal force does. As a consequence, any homogenous ellipsoid in rigid rotation around its main axis is a natural figure of equilibrium, as formulated by Maclaurin in the $18$th century \citep{chandra69}.

In this article, we investigate the toroidal analog of thin ellispoidal homoeoids. While tori, disks and rings are commonly observed over a vast range of scales in the Universe (around planets, thin and thick disks around forming and evolved stars, in galaxies), this topic has received almost no attention yet. Besides the physics of gravitation, various domains of science are concerned by toroidal structures : electrostatics \citep{Andrews2006664}, plasma physics \citep{ev81,tt00}, nuclear physics \citep{WONG1972446}, nano-structure physics and biology \citep{Kuyucak199822}.

The determination of gravitational attraction is a complex technical task, especially for toroids \citep[see e.g.][]{ct00,ba11,kt16,ma18}. Here, we focus on the potential in the cavity of a homogeneous toroidal shell with {\it a circular main radius and a circular core section}. It is shown that not only the potential but all the partial derivatives can be expressed in terms of products of elliptic integrals of the first and second kinds {\it at the pole of toroidal coordinates}. As a consequence of this exceptional property, the interior potential can be expanded as a bivariate series of $R$ and $Z$, similar to the classical Taylor expansion. We derive the leading terms and show the performance of this approach, which is particularly good for shells with a small axis ratio $e$. We find that the interior potential is roughly linear with the cylindrical radius and weakly sensitive to the altitude $Z$, in contrast with ellipsoidal homoeoids.

The article is organized as follows. In Sect. \ref{sec:focus}, we remind the potential of a massive loop and show why the pole of toroidal coordinates plays a special role for the toroidal shell with a circular core section. The exact formula for the potential at the pole is established in Sect. \ref{sec:potatpole}. The corresponding acceleration is derived in Sect. \ref{sec:gratpole}. We then generate a two-term expansion of the potential around the pole and valid in the shell cavity. This is the aim of Sect. \ref{sec:intpot}. We illustrate the method for a shell having an axis ratio of $e=0.1$. In this case, the interior potential is precise up to $4$ digits. We show in Sect. \ref{sec:higherorders} how to proceed to next orders. Two additional terms are explicitly derived. The performance of the expansion versus $e \in[0,1]$ is discussed in Sect. \ref{sec:effectofaxisratio}. A short driver program is given in Appendix. In Sect. \ref{sec:byproducts}, we determine the potential at the surface of the shell and consider the case of shells with a thick edge. We also propose an empirical law for the exterior potential which is reliable for $e^2 \ll 1$. The last section is devoted to a large discussion about the implementation of the formula, and in particular the gain with respect to a pure numerical treatment. We remind the importance of analytical solutions to understand the physics of systems hosting a massive (self-gravitating) torus. We end with a few remarks and perspectives.

\section{Loop potential and the pole}
\label{sec:focus}

\begin{figure}
\includegraphics[width=7.8cm,bb=0 0 473 263, clip=true]{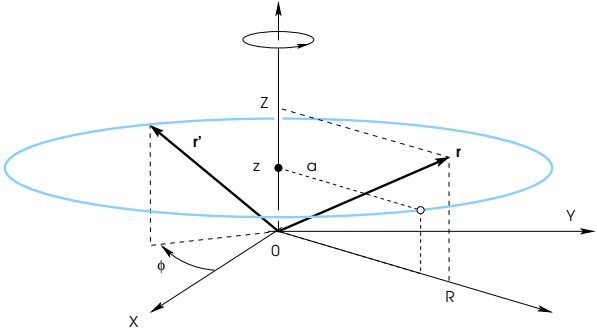}
\caption{The massive circular loop ({\it blue}). Its radius is $a$ and its center is at $(0,z)$. The potential at $\mathbfit{r}$ is given by Eq. (\ref{eq:psiloop}); see also Fig. \ref{fig:pot_wire.ps}.}
\label{fig:loop.eps}
\end{figure}

\subsection{The massive loop}

\begin{figure}
\includegraphics[width=6.8cm,bb=50 50 554 770,clip=true,angle=-90]{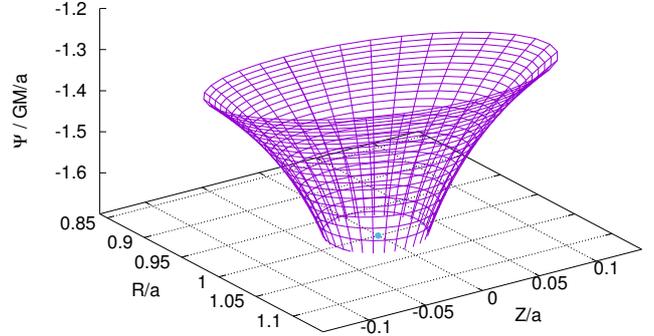}
\caption{Gravitational potential $\Psi$ of the massive loop as obtained from Eq. (\ref{eq:psiloop}). The radius of the loop is $a$ and it is positionned at $z=0$ ({\it blue dot}); see also Fig. \ref{fig:loop.eps}. The potential diverges when $R \rightarrow a$ and $Z \rightarrow z$.}
\label{fig:pot_wire.ps}
\end{figure}

We consider a massive circular loop of matter with radius $a$ and mass $2 \pi a \lambda$, as shown in Fig. \ref{fig:loop.eps}. The gravitational potential of this system is accessible at any point $\mathbfit{r}$ of space by integration of the Green function $1/|\mathbfit{r}-\mathbfit{r'}|$ over the polar angle $\phi$. In cylindrical coordinates $(R,Z)$, the formula is \citep[e.g.][]{kellogg29,durand64}
\begin{equation}
\Psi(R,Z) = -2G \lambda \sqrt{\frac{a}{R}} k \elik(k),
\label{eq:psiloop}
\end{equation}
where
\begin{equation}
\elik(k) = \int_0^{\frac{\pi}{2}}{\frac{dx}{\sqrt{1-k^2 \sin ^2 x}}}
\end{equation}
is the complete elliptic integral of the first kind,
\begin{equation}
k=\frac{2\sqrt{aR}}{\sqrt{(a+R)^2+(Z-z)^2}} \in [0,1],
\label{eq:k}
\end{equation}
is the modulus and $z$ is the altitude of the loop. This form assumes that the loop axis is confounded with the $Z$-axis. The potential is displayed versus $R$ and $Z$ in Fig. \ref{fig:pot_wire.ps}. It is logarithmically singular as $R \rightarrow a$ and $Z \rightarrow z$ (i.e. $k \rightarrow 1$) since $\elik(k)\rightarrow \ln \frac{4}{k'}$ \citep{bf54,fu16}, where $k'=\sqrt{1-k^2}$ is the complementary modulus.

\begin{figure}
\includegraphics[width=8.cm,bb=0 0 476 332, clip=true]{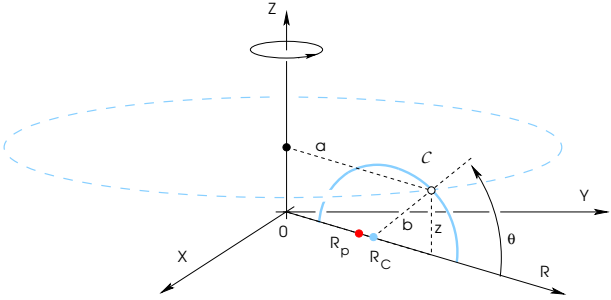}
\caption{Nominal configuration (half-plane $Z>0$ only) leading to a constant value of the modulus $k$ in Eq. (\ref{eq:k}). This situation is achieved for all points $(a,z)$ belonging to the circle ${\cal C}$ ({\it blue line}) and for $(R,Z)=(\rf,0)$, which is precisely the pole  ({\it red dot}) of the toroidal coordinates.}
\label{fig:loop2.eps}
\end{figure}

Any axially symmetrical system can be build by assembling co-axial loops, and the total potential is obtained by summing in Eq. (\ref{eq:psiloop}) over the given distribution. We easily understand that the presence of the special function renders the accounting process (be discrete or continuous) somewhat cumbersome. Hence, a question arises : are there conditions and configurations under which the modulus $k$ would be a constant in Eq. (\ref{eq:psiloop})? If we rewrite Eq. (\ref{eq:k}) in the following form,
\begin{equation}
\left[a- R\left(\frac{1+{k'}^2}{k^2}\right) \right]^2 + (z-Z)^2 = \left(\frac{2Rk'}{k^2}\right)^2,
\label{eq:kcircle}
\end{equation}
we see that points $(a,z)$ basically belong to a circle, denoted ${\cal C}$. The centre C$(\rc,\zc)$ of this circle is at
\begin{flalign}
\begin{cases}
  \rc = R \frac{1+{k'}^2}{k^2},\\
  \zc = Z,
\end{cases}
\label{eq:circleparameters}
\end{flalign}
and its radius is
\begin{flalign}
 b = R \frac{2k'}{k^2}.
\label{eq:b}
\end{flalign}
It is shown in Fig. \ref{fig:loop2.eps}. We conclude that $k$ is constant along ${\cal C}$, provided $R/\rc$ and $Z/\rc$ are fixed. We will see below which point of space is concerned. By varying $k$ in the allowed range, one gets a series of circles that are not concentric and do not have the same radius. For $k=0$, both $\rc$ and $b$ are infinite, and the circle is tangent to the vertical axis at the origin. Conversely, for $k=1$, the circle has null radius and $(R,Z)$ coincides with the centre C. 

\subsection{Link with toroidal coordinates}

\begin{figure}
\includegraphics[width=8.5cm,bb=0 0 463 406,clip=true]{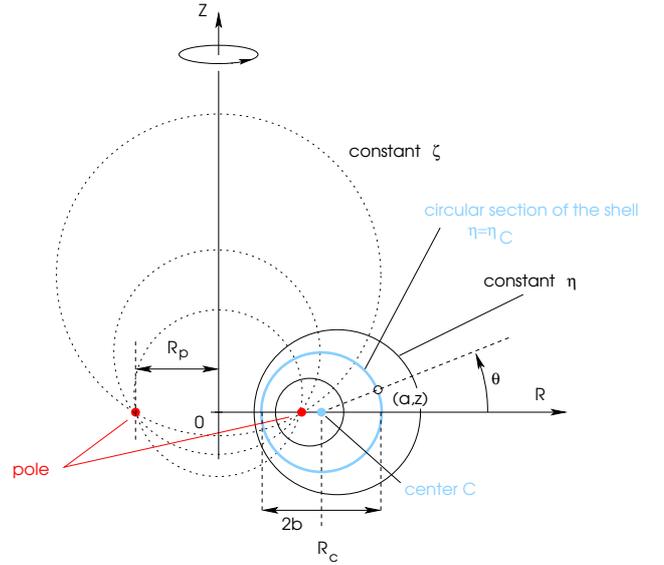}
\caption{Toroidal coordinates ($\eta,\zeta)$ shown in the $(R,Z)$ plane in the form of contour lines. Lines of constant $\eta$ ({\it plain lines}) are non-concentric circles. The shell section is one of these ({\it blue}), i.e. $\eta= \etac$. The pole (or focal ring) is at $R=\rf$ ({\it red}). At that point, $k$ is constant in Eq. (\ref{eq:k}) for any point $(a,z) \in {\cal C}$.}
\label{fig:scheme.eps}
\end{figure}

In axial symmetry, the toroidal coordinates $(\eta,\zeta)$ are linked to the cylindrical coordinates by
\begin{flalign}
\begin{cases}
 R= \rf \frac {\sinh \eta}{\cosh \eta - \cos \zeta},\\
 Z= \rf \frac {\sin \zeta}{\cosh \eta - \cos \zeta},
\end{cases}
\label{eq:rztoroidal}
\end{flalign}
where $\rf>0$ corresponds to the pole (or focal ring). This point, which belongs to the radial axis, is singular in the sense that $R \rightarrow \rf$ when  $\eta \rightarrow \infty$. By eliminating $\zeta$ in Eq. (\ref{eq:circleparameters}), we see that lines of constant $\eta$ are non-intersecting circles surrounding the pole. In terms of cylindrical coordinates $(a,z)$, the equation of a given circle is 
\begin{equation}
\left(a- \rf\coth \eta  \right)^2 + z^2 = \left(\frac{\rf}{\sinh \eta}\right)^2,
\label{eq:constanttau}
\end{equation}
where  $(\rf\coth \eta,0)$ are the cylindrical coordinates of its centre and $\rf/\sinh \eta$ is the radius. This is summarized in Fig. \ref{fig:scheme.eps}. A quick inspection of Eq. (\ref{eq:kcircle}) shows that Eq. (\ref{eq:constanttau}) is nothing but circle ${\cal C}$ (C,$b$) met above provided
\begin{flalign}
\begin{cases}
  R \equiv \rf,\\
  Z \equiv 0,\\
  k^2 = \frac{2 \sinh \eta}{\cosh \eta + \sinh \eta},\\
  \qquad {\rm or} \coth \eta  = \frac{1+{k'}^2}{k^2}.
\end{cases}
\label{eq:rzk}
\end{flalign}

In summary, the circle ${\cal C}$ (C,$b$) is a particular line of constant $\eta$, which value is denoted $\etac$ in the following (the region inside the circle corresponds to $\eta > \etac$). We can therefore state that {\it for any loop with parameters $(a,z) \in {\cal C}$, the modulus $k$ is constant at the pole of toroidal coordinates, and only at that point}.

\begin{figure}
\includegraphics[width=6.8cm,bb=50 55 554 770,clip=true,angle=-90]{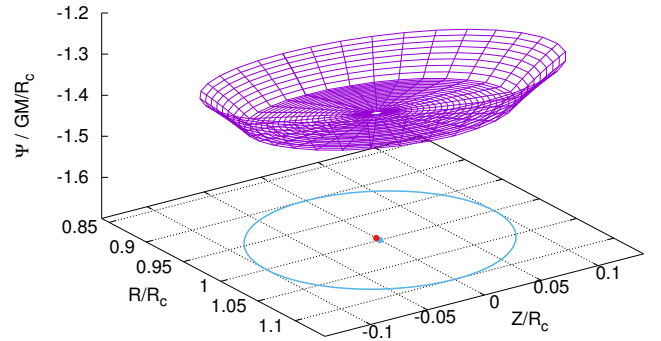}
\caption{Gravitational potential of the toroidal shell (in units of $GM/\rc$) obtained by the direct estimate of the integral in Eq. (\ref{eq:psihtorus}). The $R$- and $Z$-axis are in units of $\rc$. The normalized core radius (or axis ratio) is $b/\rc=0.1$, which corresponds to $p\approx 0.4264$, $\kc \approx 0.9987$ and $\etac \approx 2.993$. Also shown are the shell section ({\it blue line}), its center C ({\it blue dot}) and associated pole ({\it red dot}); see also Figs. \ref{fig:loop2.eps} and \ref{fig:scheme.eps}.}
\label{fig:pot.ps}
\end{figure}

\section{Potential at the pole}
\label{sec:potatpole}

\subsection{A new formula}

We now consider a continuous concatenation of co-axial massive loops, each with radius $a$ and altitude $z$, such that $(a,z) \in {\cal C}$. This forms a toroidal shell with circular section and core radius $b$ (see again Figs. \ref {fig:loop2.eps} and \ref{fig:scheme.eps}). In cylindrical coordinates, the equation of ${\cal C}$ is basically of the form
\begin{flalign}
\begin{cases}
       a=\rc + b \cos \theta,\\
     z=b \sin \theta,
\end{cases}
  \label{eq:az}
    \end{flalign}
where $\theta \in [0,2\pi]$. It follows from Eq. (\ref{eq:psiloop}) that the total potential of the shell is given by the integral
\begin{flalign}
 \Psi(R,Z) = -2G \int_0^{2\pi} { \Sigma \sqrt{\frac{a}{R}}  k \elik(k) b d\theta},
\label{eq:psihtorus}
\end{flalign}
where $\Sigma$ is the surface density\footnote{The integral bounds can be modified if, instead of a full circular section, we consider an arc.}. If all loops have the same mass per unit length, then $\Sigma$ is a constant.

An example of direct numerical integration of Eq. (\ref{eq:psihtorus}) is given in Fig. \ref{fig:pot.ps} for a homogeneous toroidal shell with axis ratio $b/\rc=0.1$, which corresponds to $\etac \sim 2.993$. As the quadrature scheme, we use the trapezoidal rule with $N_\theta=1025$ equally spaced nodes in $\theta$ (this setup is not critical). The imprint of the shell surface is clearly visible. In the exterior domain (i.e. $\eta < \etac$), the potential decreases in absolute when one moves away from the shell (i.e. $\eta \rightarrow 0$). Inside the whole cavity ($\eta > \etac$), the potential gradually increases with $R$. From the above statement (see Sect. \ref{sec:focus}), we see that $k$ (and then any function of $k$) can be taken out of the integral if $(R,Z)=(\rf,0)$. So, Eq. (\ref{eq:psihtorus}) becomes
\begin{flalign}
  \nonumber
  \psi(\rf,0) &= -2G  \frac{b}{\sqrt{\rf}} \kc \elik(\kc) \int_0^{2\pi} {\Sigma \sqrt{a} d\theta},\\
  & \equiv \psif,
\label{eq:psihtorus_out}
\end{flalign}
where $a$ is given by Eq. (\ref{eq:az}a) and $k$ has been set to $\kc$. We see the remaining integral is analytical for a wide variety of angle-dependent surface densities $\Sigma(\theta)$. The result is straightforward for the homogeneous shell. We actually have
\begin{flalign}
  \nonumber
  \int_0^{2\pi} {\sqrt{a} d\theta} & = 2\sqrt{\rc+b} \int_0^\pi{\sqrt{1-p^2 \sin ^2 \alpha}d\alpha},\\
  & = 4\sqrt{\rf} \frac{\elie(p)}{\sqrt{p'}},
\label{eq:integsqrta}
\end{flalign}
 where 
\begin{equation}
p^2=\frac{2e}{1+e}\in [0,1],
\label{eq:p}
\end{equation}
 is the ratio of the core diameter to the outer radius,
\begin{equation}
e=\frac{b}{\rc} \in [0,1],
\label{eq:e}
\end{equation}
is the axis ratio,
$p'=\sqrt{1-p^2}$ is the complementary modulus, and
\begin{equation}
\elie(k) = \int_0^{\pi/2}{\sqrt{1-k^2 \sin ^2 x}dx},
\end{equation}
is the complete elliptic integral of the second kind. So, Eq. (\ref{eq:psihtorus_out}) becomes
\begin{flalign}
  \psif &=-8G \Sigma b \kc \elik(\kc) \frac{\elie(p)}{\sqrt{p'}},
\label{eq:psihtorus2}
\end{flalign}
We are not aware that this expression has already been reported in the literature. Clearly, $\elik$ corresponds to the main curvature (radius $\rc)$ while $\elie$ is associated with the shell circular section (radius $b$). Note that, when $p \rightarrow 0$, $e \rightarrow 0$ while $\rf \rightarrow \rc$. In this case, the shell tends to the loop, Eq. (\ref{eq:psihtorus2}) matches Eq. (\ref{eq:psiloop}) and its mass converges to $2 \pi \lambda a$. At the opposite, for $p \rightarrow 1$, the shell has infinite radius, mass and potential.

\subsection{Equivalent forms}

\begin{figure}
\includegraphics[height=8.7cm,bb=50 50 554 770, angle=-90, clip=true]{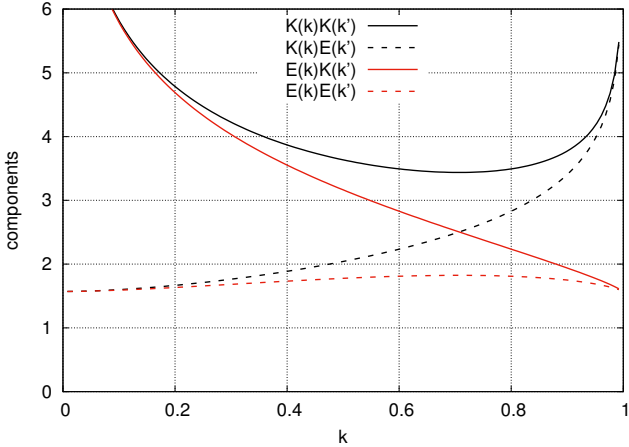}
\caption{The four components of the basis vector $\bcal$ defined by Eq. (\ref{eq:VecB}) versus the modulus $k$.}
\label{fig:basis.ps}
\end{figure}

From Eqs. (\ref{eq:circleparameters}a), (\ref{eq:b}), (\ref{eq:p}) and (\ref{eq:e}), we see that $p$ is linked to $\kc$ through the following relations
\begin{equation}
p = \frac{2\sqrt{\kpc}}{1+\kpc},
\end{equation}
and
\begin{equation}
p' = \frac{1-\kpc}{1+\kpc},
\end{equation}
meaning that Eq. (\ref{eq:psihtorus2}) can take equivalent forms by considering modulus transformations (see the Appendix A). From Eq. (\ref{eq:kk2kprim}), we can eliminate $\kc$, and we find
\begin{equation}
\psif =-16 G \Sigma b \elik(p') \elie(p).
\label{eq:psihtorus5}
\end{equation}
Since the mass of the toroidal shell is
\begin{equation}
    M= 4 \pi^2 \Sigma b \rc,
\label{eq:mass}
\end{equation}
we can easily rewrite the potential as a function of $GM/R_c$ (or even $GM/\rf$). This is for instance
\begin{equation}
  \psif = - \frac{GM}{\rc} \frac{4}{\pi^2} \elik\left(\sqrt{\frac{1-e}{1+e}}\right) \elie\left(\sqrt{\frac{2e}{1+e}}\right).
\end{equation}
If we can eliminate $\elie(p)$ in Eq. (\ref{eq:psihtorus2}) by using Eq. (\ref{eq:ep2kprim}), we find
\begin{equation}
\psif = -8 G \Sigma b \elik(\kc) \left[ 2\elie(\kpc)-\kc^2 \elik(\kpc)\right].
\label{eq:psihtorus4}
\end{equation}
In order to anticipate a little bit, we introduce the four-vector
\begin{equation}
  \bcal(k)= 
  \begin{pmatrix}
    \elik(k)\elik(k')\\
    \elik(k)\elie(k')\\
    \elie(k)\elik(k')\\
    \elie(k)\elie(k')
  \end{pmatrix},
  \label{eq:VecB}
\end{equation}
whose components are plotted versus $k$ in Fig. \ref{fig:basis.ps}. We see that Eq. (\ref{eq:psihtorus4}) reads
\begin{equation}
\psif = - G \Sigma b \; \acal_{00}(\kc) \cdot \bcal(\kc),
\label{eq:psihtorus_AB}
\end{equation}
where
\begin{equation}
         \acal_{00}(k)=8
        \begin{pmatrix}
          -k^2\\
          2\\
          0\\
          0
        \end{pmatrix}.
\label{eq:VecA00}
\end{equation}
Note that we can expand the complete elliptic integrals as a function of $k$ in extreme regimes where $k \rightarrow 0$ and $k \rightarrow 1$. Then $\psif$ is only function of $\kc$ and $\ln \kpc$. 

\section{Acceleration at the pole}
\label{sec:gratpole}

We can make the same analysis for the gravitational acceleration. Since the pole belongs to the $R$-axis, the vertical acceleration is zero by symmetry. The formula for the radial acceleration caused by the loop can be determined from Eq. (\ref{eq:psiloop}). The classical form is \citep{durand64,hure05}
\begin{equation}
- \frac{\partial \Psi}{\partial R} = G \lambda k \sqrt{\frac{a}{R^3}}\left[  \left(1+\frac{a-R}{2a} \frac{k^2}{{k'}^2}\right)\elie(k)-\elik(k) \right].
 \label{eq:ghtorus}
\end{equation}
To get the acceleration due to the toroidal shell, we have to multiply this expression by $\Sigma b d\theta/\lambda$, and to integrate over $\theta \in [0,2\pi]$. At the pole, any function that depends only on $k$ can be carried out of the integral, which requires to set $k=\kc$ (see Sect. \ref{sec:focus}). We then find, again assuming $\Sigma$ constant :
\begin{flalign}
  \nonumber
  - \left. \frac{\partial \Psi}{\partial R} \right|_{\rf} &= \frac{G\Sigma \kc}{\sqrt{\rf^3}} \left[  \left(\frac{1+{\kpc}^2}{2{\kpc}^2}\right)\elie(\kc)-\elik(\kc) \right] b \int {\sqrt{a} d\theta}\\
  & \qquad - \frac{G\Sigma}{\sqrt{\rf}} \frac{\kc^3}{2{\kpc}^2}\elie(\kc) b \int {\frac{d\theta}{ \sqrt{a}}},
 \label{eq:ghtorus2}
\end{flalign}
The first integral has already been met; see Eq. (\ref{eq:integsqrta}). The second one can be easily calculated. We find
\begin{flalign}
  \nonumber
  \int_0^{2\pi} {\frac{d\theta}{ \sqrt{a}}} &= \frac{2}{\sqrt{\rc+a'}} \int_0^\pi{\frac{d\alpha}{\sqrt{1-p^2 \sin ^2 \alpha}}},\\
  &=  \frac{4}{\sqrt{\rf}} \sqrt{p'} \elik(p) 
\label{eq:integ1oversqrta}
\end{flalign}
and so Eq. (\ref{eq:ghtorus2}) becomes
\begin{flalign}
  \nonumber
  - \left. \frac{\partial \Psi}{\partial R} \right|_{\rf}   = \frac{4G\kc \Sigma b}{\rf}  & \left\{ \left[  \left(\frac{1+{\kpc}^2}{2{\kpc}^2}\right)\elie(\kc)-\elik(\kc) \right] \frac{\elie(p)}{\sqrt{p'}} \right.\\
  & \qquad \left. - \frac{\kc^2}{2{\kpc}^2}\elie(\kc) \sqrt{p'} \elik(p) \right\}.
 \label{eq:ghtorus3}
\end{flalign}

By using Eqs. (\ref{eq:ep2kprim}) and (\ref{eq:kp2kkprim}), we can eliminate both $\elik(p)$ and $\elie(p)$. The final expression then depends only on $\elie$ and $\elik$ evaluated at $\kc$ and $\kpc$, and it can be put in a form similar as Eq. (\ref{eq:psihtorus_AB}), namely
\begin{equation}
\left. \frac{\partial \Psi}{\partial R} \right|_{\rf} = - \frac{G\Sigma b}{\rf} \acal_{10}(\kc) \cdot \bcal(\kc),
\label{eq:ghtorus4}
\end{equation}
where 
\begin{equation}
\acal_{10}(k)=\frac{4}{k'^2}
\begin{pmatrix}
  k^2k'^2\\
  -2k'^2\\
  -k^2\\
  1+k'^2
\end{pmatrix}.
\label{eq:VecA10}
\end{equation}
As done for the potential, we can write this term as a function of the mass of the shell by using Eq. (\ref{eq:mass}).

\section{Route to the interior potential}
\label{sec:intpot}

\begin{figure}
  \includegraphics[height=8.7cm,bb=50 50 554 770,clip=true,angle=-90]{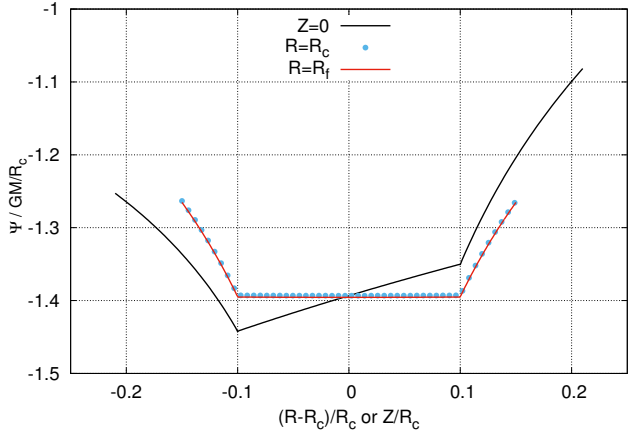}
\caption{Gravitational potential of the toroidal shell in units of $GM/\rc$ at the equatorial plane ({\it black line}), at $R=R_f$ ({\it red line}) and at $R=R_c$  ({\it blue dots}). The axis are in units of $\rc$. The conditions are the same as for Fig. \ref{fig:pot.ps}.}
\label{fig:pot_3axis.ps}
\end{figure}

There are three striking properties of $\Psi$ inside the cavity of the toroidal shell where $\eta \ge \etac$. First, the impact of variable $Z$ is very weak (almost not visible by eyes), especially if the core radius is small. Second, $\Psi$ is an increasing function of $R$ as already mentioned. The gravitational acceleration is oriented towards the origin of coordinates for $R \ge \rc -b$. Third, the variation of $\Psi$ with $R$ is very close to linear. Proofs are given in Fig. \ref{fig:pot_3axis.ps}, which shows $\Psi$ at the equatorial plane (i.e. for $Z=0$), at $R=\rf$ and at $R=R_c$ for the shell considered in Fig. \ref{fig:pot.ps}. The contrast with the ellipsoidal shell is therefore evident: the {\it interior potential is not a constant but has a small positive gradient with $R$} due to the curvature around the $Z$-axis \citep[e.g.][]{kellogg29,binneytremaine87}. Actually, for an observer standing at the inner edge of the shell and looking towards the origin, matter is present at relatively short seperations behind (the outer edge) and in front (opposite inner and outer edges). In contrast, at the outer edge, there is no matter behind and separations are larger. The potential well is therefore deeper at the inner edge, which is rather intuitive.

It is tempting to elaborate some kind of a fit. We could produce sets of data by varying $R/\rc$, $Z/\rc$ and $e$, but it seems more powerful to consider a bivariate series, resembling the Taylor series, i.e.
\begin{flalign}
   \label{eq:bivtaylor}
 & \Psi(R,Z) = \psif +  (R-\rf) \left. \frac{\partial \Psi }{\partial R}\right|_\rf + Z \left. \frac{\partial \Psi }{\partial Z}\right|_\rf  \\
  & \qquad + \frac{1}{2}(R-\rf)^2 \left. \frac{\partial^2 \Psi}{\partial R^2}\right|_\rf + (R-\rf)Z \left. \frac{\partial^2 \Psi}{\partial Z \partial R}\right|_\rf  + \dots.
 \nonumber,
\end{flalign}
since the first terms have already been calculated. Any expansion is, however, necessarily limited to the cavity and cannot be valid in the whole physical space. Actually, $\Psi$ is continuous but not differenciable at the surface of the shell. Besides, the potential must satifiy the Laplace equation in the cavity, namely
\begin{equation}
  \frac{\partial^2 \Psi}{\partial R^2}  + \frac{1}{R}\frac{\partial \Psi}{\partial R} + \frac{\partial^2 \Psi}{\partial Z^2} =0,
  \label{eq:laplace}
\end{equation}
which is a priori not automatic with the above form.

\subsection{Zero-order approximation}

We start with the crudest approximation, i.e.
\begin{equation}
  \Psi(R,Z) \approx C_0 = {\rm const.},
  \label{eq:ax0pot}
\end{equation}
in which case Eq. (\ref{eq:laplace}) is trivially fulfilled. By constant, we mean that the potential is uniform inside the cavity. Obviously, this constant value is function of the shell parameters $\Sigma$, $b$ and $\kc$ (or $e$). We then have $C_0 = \psif$. We have checked this approximation by comparing $\psif$ to values  (hereafter ``reference values'') obtained by the direct integration of Eq. ({\ref{eq:psihtorus}); see Sect. \ref{sec:potatpole}. The conditions are the same as for Fig. \ref{fig:pot.ps}. Data have been produced in the cavity only, which is easily done in toroidal coordinates by varying $\eta$ in the range $[\etac,\infty]$ and $\xi \in [0,2\pi]$. The logarithmic deviation is shown in Fig. \ref{fig:err0.ps} as a function of $R/\rc$ and $Z/\rc$. As expected, the potential given by Eq. (\ref{eq:ax0pot}) is underestimated close to the inner edge, and overestimated close to the outer edge. We see, however, that the agreement is globally correct and even better than expected. The mean relative deviation is $\sim 10^{-2.23}$, i.e. less than one percent. We see that the precision is nominal not only around the pole but also along two directions $|\theta| \sim \pi/2$. If the core radius is decreased, the acceleration also decreases, the interior potential is flatter, and the approximation becomes even better (see Sect. \ref{sec:effectofaxisratio}). 

\begin{figure}
  \includegraphics[width=8.6cm,bb= 56 265 552 695,clip=true]{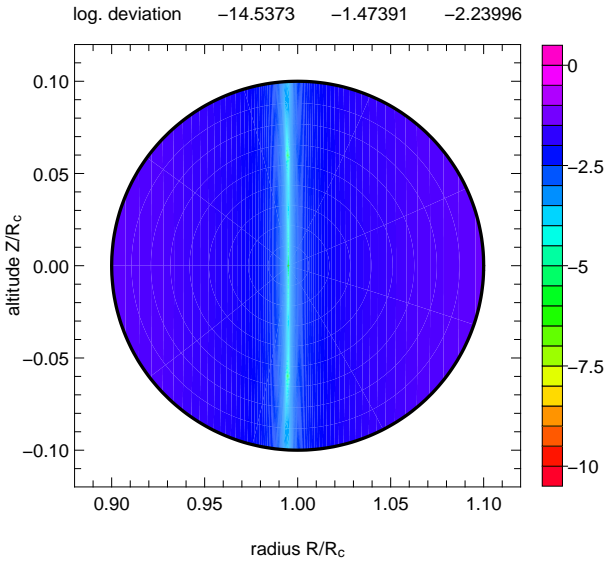}
\caption{The decimal logarithm of the relative error between  $\Psi$ estimated by direct integration and $\psif$ (numbers at the top, from left to right, refer to the min., max. and mean values respectively). The shell is the same as for Fig. \ref{fig:pot.ps}, i.e. $e=0.1$. The axis are in units of $\rc$. The computations are restricted to the interior cavity where $\eta \ge \etac$.}
\label{fig:err0.ps}
\end{figure}

\subsection{First-order approximation (the curvature effect included)}

In the absence of any variation in $Z$, the radial contribution must vanish in Eq. (\ref{eq:laplace}), which means
\begin{equation}
  \Psi(R,Z) \approx C_0 +C_1 \ln R,
  \label{eq:ax1pot}
\end{equation}
where $C_0$ and $C_1$ are to be determined. In particular, at the pole, we have
\begin{flalign}
  \begin{cases}
    \psif = C_0 + C_1\ln \rf, \\
    \gf = -\frac{C_1}{\rf}.
   \label{eq:ab_order1}
   \end{cases}
\end{flalign}
Since $\psif$ and $\gf$ are known (see Sect. \ref{sec:potatpole} and \ref{sec:gratpole}), this set of equations is easily solved for $C_0$ anf $C_1$, making Eq. (\ref{eq:ax1pot}) fully operational. We have compared this new approximation to reference values under the same conditions as above. Figure \ref{fig:err1.ps} displays the decimal logarithm of the relative error. We notice that the precision is improved by more than two orders of magnitude, with a mean deviation of $\sim 10^{-4.70}$. It is nominal around the pole, as well as in four directions $|\theta| \sim \pi/4$ and $3\pi/4$. Note that the quasi-linear behavior of $\Psi$ in the cavity is especially marked for shells with a small axis ratio. This is explained by expanding $\ln R$ around $\rf$, which gives
\begin{flalign}
\Psi \approx  \psif +  C_1 \frac{R-\rf}{\rf}.
   \label{eq:quasilinear}
\end{flalign}

\begin{figure}
  \includegraphics[width=8.6cm,bb=56 265 552 695,clip=true]{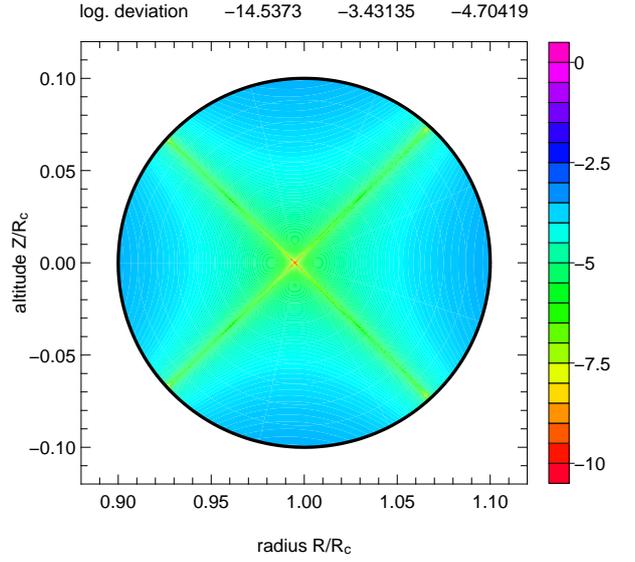}
\caption{The legend is the same as for Fig. \ref{fig:err0.ps}, but the approximation given by Eq. (\ref{eq:ax1pot}).}
\label{fig:err1.ps}
\end{figure}

\section{Next orders}
\label{sec:higherorders}

\subsection{Second-order approximation}
\label{subsec:order2}

The variation of $\Psi$ inside the cavity is much more complex, with a sensitivity to variable $Z$ (in fact, to $Z^2$ due to equatorial symmetry). If we allow for quadratic contributions, the vacuum solution is of the form
\begin{equation}
\Psi(R,Z) \approx C_0 + C_1 \ln R + C_2\left( R^2 -2 Z^2\right),
\label{eq:ax2pot}
\end{equation}
where $C_0$, $C_1$ and $C_2$ are three quantities to be determined$^2$. At the pole, we have
\begin{flalign}
  \begin{cases}
    \psif = C_0 + C_1 \ln \rf + C_2 \rf^2,\\
   \gf = -\frac{C_1}{\rf} - 2C_2 \rf,
   \label{eq:ab_order2}
   \end{cases}
\end{flalign}
and we see that a third equation is needed to fix $C_2$. This is done by calculating the vertical gradient of the vertical acceleration, namely
\begin{equation}
  4C_2 = -\left.\frac{\partial^2 \psi}{\partial Z^2}\right|_\rf.
  \label{eq:c}
\end{equation}
Alternatively, we could also consider $\partial^2_{R^2} \Psi$ (since $\Psi$ is harmonic, derivatives with respect to $Z$ and $R$ are linked together; see below). We then reiterate the procedure described in Sects. \ref{sec:potatpole} and \ref{sec:gratpole}. The vertical acceleration due to the circular loop is given by \citep{durand64,hure05}
\begin{equation}
  - \frac{\partial \psi}{\partial Z} = \frac{G \lambda(z-Z)}{2\sqrt{aR^3}} \frac{k^3\elie(k)}{{k'}^2},
  \label{eq:gzring}
\end{equation}
and so the second derivative writes
  \begin{flalign}
    \label{eq:d2psioverdz2}
    - \frac{\partial^2 \psi}{\partial Z^2} = & \frac{G \lambda}{2\sqrt{aR^3}} \left\{-\frac{k^3\elie(k)}{{k'}^2} \right.\\
    & \left. \qquad + (z-Z) \frac{\partial}{\partial k} \left[\frac{k^3\elie(k)}{{k'}^2} \right] \frac{\partial k}{\partial Z} \right\}.
  \nonumber
  \end{flalign}
From Eq. (\ref{eq:k}), we find
\begin{equation}
    (z-Z)\frac{\partial k}{\partial Z} = (z-Z)^2\frac{k^3}{4aR}.
  \label{eq:zdkdz}
\end{equation}
Besides, from Eq. (\ref{eq:deoveerdk}) of Appendix B, we get
\begin{equation}
    \frac{\partial}{\partial k} \left[\frac{k^3\elie(k)}{{k'}^2} \right] = \frac{k^2}{{k'}^4} \elil(k), 
    \label{eq:dk3eoverkprim2dz}
  \end{equation}
where we have defined
\begin{equation}
  \elil(k) =  2(1+{k'}^2) \elie(k) -{k'}^2 \elik(k),
\end{equation}
for convenience. To get $\partial^2_{Z^2} \Psi$ for the toroidal shell, we just have to multiply Eq. (\ref{eq:d2psioverdz2}) by $\Sigma bd\theta/\lambda$ and to integrate over $\theta$. In this process, $a$ and $z$ are still given by Eq. (\ref{eq:az}). The complexity of the calculus is apparent since, at the pole, $k$ is a constant (see Sect. \ref{sec:focus}). Only terms with $a$ and $z$ are therefore retained in the integral, and we set $k=\kc$. We then have
\begin{flalign}
  -  \left. \frac{\partial^2 \psi}{\partial Z^2}\right|_\rf  = \frac{Gb}{2\sqrt{\rf^3}} & \left\{ - \frac{\kc^3\elie(\kc)}{{\kpc}^2}\int{\frac{\Sigma d \theta}{\sqrt{a}}} \right. \\
  & \qquad \left. +  \frac{{b}^2}{4\rf} \frac{\kc^5}{{\kpc}^4} \elil(k) \int{\frac{\Sigma \sin^2 \theta d \theta}{a\sqrt{a}}} \right\},
  \nonumber
\end{flalign}
where the term $\sin^2 \theta$ comes from $(z-Z)^2$, evaluated at the pole. The first integral has already been employed above, and the second one is analytical, namely
\begin{flalign}
\label{eq:integsin2overasqrta}
   \int_0^{2\pi}{\frac{\sin^2 \theta d \theta}{a\sqrt{a}}} & = \frac{8}{\sqrt{\rc+a'}^3}\int_0^\pi{\frac{\cos ^2 \alpha \sin ^2 \alpha d\alpha}{\sqrt{1-p^2 \sin ^2 \alpha}^3}}\\
  \nonumber
  & = \frac{16 }{\sqrt{\rf^3}} \frac{\sqrt{{p'}^3}}{p^4}  \left[ (1+{p'}^2) \elik(p) - 2 \elie(p) \right].
\end{flalign}

After some algebra and tedious calculus, we find
\begin{flalign}
     \label{eq:d2psioverdz2bis}
 -   \left. \frac{\partial^2 \psi}{\partial Z^2}\right|_\rf  & =  \frac{2Gb\kc^3 }{\rf^2 {\kpc}^2}   \left\{ - \elie(\kc)\sqrt{p'} \elik(p) \right. \\
 \nonumber
  &  \qquad \left. + \frac{\kc^2}{4{\kpc}^2\sqrt{p'}} \elil(\kc)  \left[(1+{p'}^2) \elik(p) - 2 \elie(p)\right] \right\},
\end{flalign}
which leads to $C_2$ from Eq. (\ref{eq:c}), and subsequently to $C_0$ and $C_1$ from Eq. (\ref{eq:ab_order2}). As done before, we can express $\elik(p)$ and $\elie(p)$ as functions of $\elik(\kpc)$ and $\elie(\kpc)$, and put the result in the form of a scalar product. The final formula is
\begin{equation}
     \label{eq:d2psioverdz2bis}
   \left. \frac{\partial^2 \psi}{\partial Z^2}\right|_\rf  = - \frac{G\Sigma b}{\rf^2} \acal_{02}(\kc) \cdot \bcal(\kc),
\end{equation}
where
\begin{equation}
  \acal_{02}(k)=2\frac{k^4}{k'^4}
\begin{pmatrix}
  -k'^2\\
  k'^2\\
  2+k'^2\\
  -2-2k'^2
  \label{eq:VecA02}
  \end{pmatrix},
\end{equation}
while $\bcal$ is given by the Eq.  (\ref{eq:VecB}). Again, this derivative can be expressed as a function of the mass of the shell and $R_c$ or $\rf$. From Eqs. (\ref{eq:ghtorus4}), (\ref{eq:laplace}) and (\ref{eq:d2psioverdz2bis}), we deduce
\begin{equation}
     \label{eq:d2psioverdr2}
   \left. \frac{\partial^2 \psi}{\partial R^2}\right|_\rf  = - \frac{G\Sigma b}{\rf^2} \acal_{20}(\kc) \cdot \bcal(\kc),
\end{equation}
where
\begin{equation}
\acal_{20}(k) = - \acal_{02}(k) - \acal_{10}(k).
\end{equation}

Figure \ref{fig:err2.ps} displays the decimal logarithm of the relative error between Eq. (\ref{eq:ax2pot}) and reference values. The numerical setup is unchanged. We see that the deviation is now reduced by almost two orders of magnitude, with a mean value of $\sim 10^{-6.47}$. The precision is nominal in the vicinity of the pole, as well as in now six directions $|\theta| \sim \pi/6, 3\pi/6$ and $4\pi/6$.

\begin{figure}
  \includegraphics[width=8.6cm,bb=56 265 552 695,clip=true]{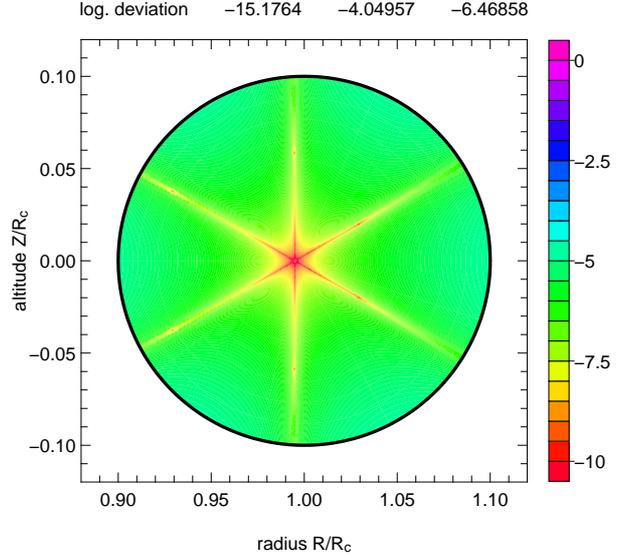}
\caption{The legend is the same as for Fig. \ref{fig:err0.ps}, but the approximation given by Eq. (\ref{eq:ax2pot}).}
\label{fig:err2.ps}
\end{figure}

\subsection{Third-order approximation}
\label{subsec:order4}

We can in principle continue the process up to the desired order, say $N+1 \ge 2$. The expansion is then supplemented with a new set of terms
\begin{equation}
  C_{2N} \left( R^{2N} +\sum_{i=1}^{N} \alpha_{i,N-i} R^{2N-2i} Z^{2i}\right),
\end{equation}
where the coefficients $\alpha_{i,N-i}$ are found from Eq. (\ref{eq:laplace}), and the leading quantity $C_{2N}$ is determined from a $N$th-order partial derivative of the potential evaluated at the pole, which can be put into the form
\begin{equation}
\left. \frac{\partial^{n+m} \Psi}{\partial R^n Z^m} \right|_{\rf} = - \frac{G\Sigma b}{\rf^{n+m}} \acal_{nm}(\kc) \cdot \bcal(\kc),
\end{equation}
where $N+1=m+n$, where $m \ge 0$, $n \ge 0$, and $\acal_{nm}(k)$ is a four-vector whose components are rational functions of $k$. This is the most tricky part of the method. Here are the results for $N=2$. Including quadratic terms, the vacuum potential has the form
\begin{flalign}
\label{eq:ax4pot}
  \psi(R,Z) \approx C_0&+ C_1 \ln R + C_2\left( R^2 -2 Z^2\right)\\
  &\qquad +  C_4 \left(R^4-8R^2Z^2+\frac{8}{3}Z^4 \right),
  \nonumber
\end{flalign}
where there are now four unknowns involved. At the pole, we have
\begin{flalign}
  \begin{cases}
    \psif = C_0 + C_1 \ln \rf + C_2 \rf^2 +C_4 \rf^4\\
    \gf = -\frac{C_1}{\rf} - 2C_2 \rf -4 C_4 \rf^3, \\
    - \left. \frac{\partial^2 \psi}{\partial Z^2}\right|_\rf = 4C_2 + 16 C_4 \rf^2
   \label{eq:abcorder4}
   \end{cases}
\end{flalign}
which can be solved if we can add a new equation. We see that $C_4$ is accessible from a cross-derivative, namely
\begin{equation}
  - 32 C_4 \rf = \left.\frac{\partial^3 \psi}{\partial R \partial Z^2}\right|_\rf,
  \label{eq:dorder4}
\end{equation}
which can be calculated from Eq. (\ref{eq:d2psioverdz2}). We actually have
 \begin{flalign}
   \nonumber
   - \frac{\partial^3 \psi}{\partial R \partial Z^2} & = \frac{3G\lambda}{4\sqrt{aR^5}} \frac{k^3\elie(k)}{{k'}^2} - \frac{G\lambda}{2\sqrt{aR^3}} \frac{\partial}{\partial k} \left[\frac{k^3\elie(k)}{{k'}^2} \right] \frac{\partial k}{\partial R}\\
   \nonumber
   & \qquad -  \frac{5G\lambda \xi^2}{16 \sqrt{a^3R^7}}  \frac{k^5}{{k'}^4} \elil(k) \\
   &  \qquad \qquad + \frac{G\lambda \xi^2}{8\sqrt{a^3R^5}} \frac{\partial}{\partial k} \left[\frac{k^5 \elil(k)}{{k'}^4}\right] \frac{\partial k}{\partial R}.
 \end{flalign}

From Eq. (\ref{eq:k}), we get
\begin{equation}
    \frac{\partial k}{\partial R} = \frac{k^3}{4R}\left( \frac{1+{k'^2}}{k^2}-\frac{R}{a}\right),
  \label{eq:zdkdz}
\end{equation}
and we find (see also Appendix B)
 \begin{flalign}
 \frac{\partial}{\partial k} \left[\frac{k^5 \elil(k)}{{k'}^4}\right] = \frac{k^4}{{k'}^6} \left[4(1+{k'}^2) \elil(k) -9 {k'}^2 \elie(k)\right],
 \end{flalign}
 with the help of Eqs. (\ref{eq:deoveerdk}) and (\ref{eq:dkoveerdk}). At the pole, we set $R=\rf$, $Z=0$, and $k=\kc$. We see that two integrals involving $1/\sqrt{a^3}$ and $z^2/\sqrt{a^5}$ appear. Still assuming $\Sigma=$const., we have
\begin{flalign}
  \nonumber
  \int_0^{2\pi} {\frac{d\theta}{ a\sqrt{a}}} &= \frac{2}{\sqrt{\rc+a'}^3} \int_0^\pi{\frac{d\alpha}{\sqrt{1-p^2 \sin ^2 \alpha}^3}},\\
  &=  \frac{4}{\sqrt{\rf^3}} \frac{\elie(p)}{\sqrt{p'}},
\label{eq:integ1oversqrta3}
\end{flalign}
and
\begin{flalign}
\label{eq:integsin2overa2sqrta}
  \int_0^{2\pi}{\frac{\sin^2 \theta d \theta}{a^2\sqrt{a}}} & = \frac{8}{\sqrt{\rc+a'}^5}\int_0^\pi{\frac{\cos ^2 \alpha \sin ^2 \alpha d\alpha}{\sqrt{1-p^2 \sin ^2 \alpha}^5}}\\
  & = \frac{16 \sqrt{{p'}^5}}{\sqrt{\rf^5}} \frac{1}{3p^4{p'}^2} \left[ \elie(p) (1+{p'}^2)-2 {p'}^2 \elik(p) \right],
  \nonumber
\end{flalign}
respectiveley. So, the right-hand-side of Eq. (\ref{eq:dorder4}) is
\begin{equation}
 \left.\frac{\partial^3 \psi}{\partial R \partial Z^2}\right|_\rf  =  - \frac{G\Sigma b}{\rf^3} \acal_{12}(\kc) \cdot \bcal(\kc)
\label{eq:dPdRdZ2}
\end{equation}
where we have applied, as above, modulus transformations, and
\begin{equation}
  \acal_{12}(k)=\frac{2k^4}{3k'^6} 
\begin{pmatrix}
 k'^2 (-4+3k'^2-2k'^4)\\
  k'^2(4-k'^2+4k'^4)\\
  8-7k'^2-2k'^4+4k'^6\\
  (1+k'^2)(-8+11k'^2-8k'^4)
  \end{pmatrix}.
\label{eq:Vec21}
\end{equation}
The $3$rd-order approximation is then ready to be used since $C_0$, $C_1$, $C_2$ and $C_4$ are found from Eqs. (\ref{eq:abcorder4}) and (\ref{eq:dorder4}). The reader can find in the Appendix \ref{sec:f90program} a short driver program written in Fortran 90 to compute $\Psi$ from Eq.(\ref{eq:ax4pot}). The comparison of this new approximation with direct numerical integration is shown in Fig. \ref{fig:err4.ps}. The conditions are the same as above. The precision is increased compared to $2$nd-order, by two more digits typically. In the cavity, the mean deviation is now $\sim 10^{-8.00}$. The pattern is similar : the smallest deviations are found around the pole and along eight directions defined by $|\theta| \sim \pi/8, 3\pi/8, 5\pi/8$ and $7\pi/8$.

\begin{figure}
  \includegraphics[width=8.6cm,bb=56 265 552 695,clip=true]{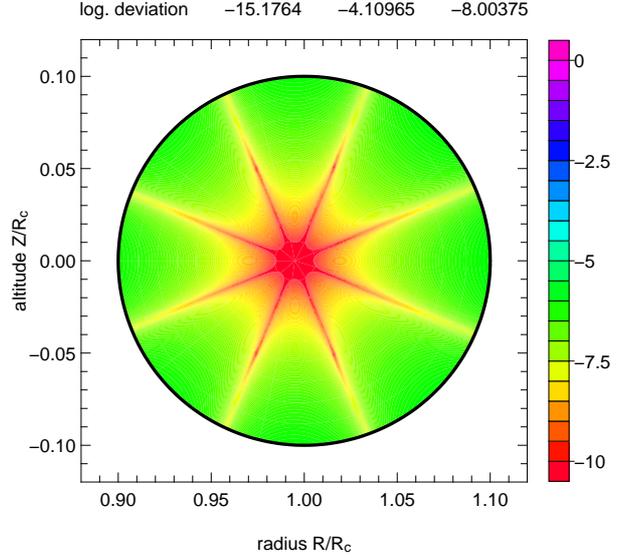}
\caption{The legend is the same as for Fig. \ref{fig:err0.ps}, but the approximation is given by Eq. (\ref{eq:ax4pot}).}
\label{fig:err4.ps}
\end{figure}
 
\begin{figure}
\includegraphics[height=8.6cm,bb=50 55 554 770,clip=true,angle=-90]{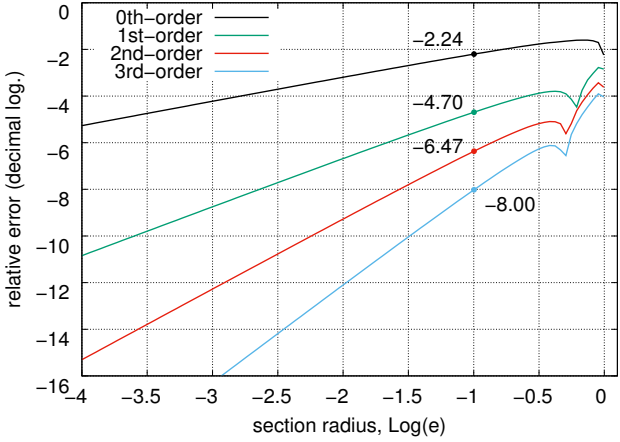}
\caption{Mean deviations between the direct numerical integration of Eq. (\ref{eq:psihtorus}) and the four approximations versus the axis ratio $e$ of the shell :  zero-order is for Eq. (\ref{eq:ax0pot}), first-order for Eq. (\ref{eq:ax1pot}), second-order for Eq. (\ref{eq:ax2pot}) and third-order for Eq. (\ref{eq:ax4pot}).}
\label{fig:errvsaprim.ps}
\end{figure}

\section{Effect of the shell axis-ratio}
\label{sec:effectofaxisratio}

Examples given above concern a toroidal shell with $e=0.1$ (i.e. $\kc \approx 0.9987$). It is interesting to see how all these expansions behave when varying the axis ratio. We repeated the computations for the full range, i.e. $e \in \, ]0,1[$. For each parameter, potential values inside the cavity have been generated by direct computation of Eq. (\ref{eq:psihtorus}), and for the four approximations, i.e. from Eq. (\ref{eq:ax1pot}) for order $0$,  Eq. (\ref{eq:ax1pot}) for order $1$,  Eq. (\ref{eq:ax2pot}) for order $2$ and  Eq. (\ref{eq:ax4pot}) for order $3$. The log. of the relative differences, i.e. $\log |\Delta \Psi/\Psi|$, have been averaged. The results are displayed in Fig. \ref{fig:errvsaprim.ps}. Unsurprisingly, none of the approximations is really reliable for the largest values of $e$. We remind that, when $b \rightarrow \rc$, the inner edge of the shell is close to the origin, while its outer edge tends to infinity, and curvature effects are important at short radii. In fact, the interior potential shows a complex, saddle-type structure. In contrast, as soon as $e \lesssim 0.3$, the four approximations are very efficient. Accuracy increases as the shell section gradually decreases. The sensitivity (i.e. slope) depends on the order too. We find
  \begin{flalign}
    \langle \log \left|\frac{\Delta \Psi}{\Psi} \right|\rangle  \sim
    \begin{cases}
   -1.41 + 0.89  \log e \qquad {\rm at \; order \; 0}, \\
   -3.03 + 1.81 \log e \qquad {\rm at \; order \; 1}, \\
   -3.79 + 2.77 \log e \qquad {\rm at \; order \; 2}, \\
   -4.28 + 3.86 \log e \qquad {\rm at \; order \; 3}.
   \label{eq:errvsorders}
   \end{cases}
\end{flalign}

\section{Byproducts}
\label{sec:byproducts}

\begin{figure}
\includegraphics[height=8.7cm,bb=0 50 554 770,clip=true,angle=-90]{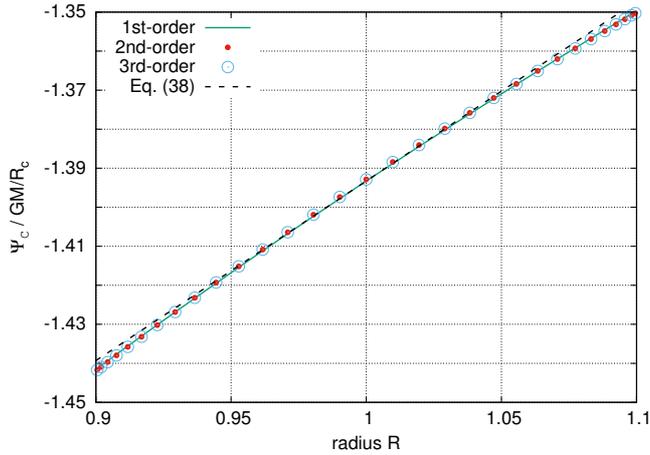}
\caption{Gravitational potential $\Psigamma$ in units of $GM/\rc$ along the section of the shell as estimated from the linear approximation, i.e. Eq. (\ref{eq:quasilinear}), and from Eqs. (\ref{eq:psigamma})b-d.}
\label{fig:pot_gamma.ps}
\end{figure}

\subsection{Potential at the surface of the shell}
\label{subsec:potsurf}

Potential values on the shell (i.e. along the circle ${\cal C}$) can be estimated from any approximation derived above since the potential is continuous inside the cavity. These values can be used for instance as approximate (but accuracte) Dirichlet boundary conditions for solving the Poisson equation. There are different equivalent options, depending on the variable selected to describe the section of the shell, mainly $R(Z)$, $Z(R)$, $\theta$ or $\eta$, which can even be converted into $k$ from Eq. (\ref{eq:rzk}c). If we use cylindrical variables, then $Z(R)^2=b^2-(R-\rc)^2$ and so we get
\begin{flalign}
  \Psigamma \approx
  \begin{cases}
    C_0 \qquad \text{at order 0},\\\\
    C_0 + C_1\ln R \qquad \text{at order 1},\\\\
    C_0 + C_1\ln R+C_2(3R^2 +2\rc^2-2b^2-4R R_0)\\
    \qquad \text{at order 2},\\\\
    C_0 + C_1\ln R+\frac{35}{3}C_4R^4-\frac{80}{3}C_4 \rc R^3\\
    \quad + \left(24 C_4 \rc^2 - \frac{40}{3} C_4 b^2 + 3C_2 \right) R^2\\
    \quad+ \left(-\frac{32}{3}C_4\rc^3+\frac{32}{3}b^2 \rc -4 C_2 \rc \right)R\\
    \quad + \frac{8}{3}C_4(\rc^4-2 C_1^2 \rc^2)\\
    \qquad \text{at order 3}.
  \end{cases}
  \label{eq:psigamma}
\end{flalign}
where $R \in [\rc-b,\rc+b]$. Figure \ref{fig:pot_gamma.ps} shows the results obtained at orders $1$ to $3$ (order 0 is too crude to be retained). We see that the linear approximation given by Eq. (\ref{eq:quasilinear}) is quite acceptable. By varying $e$, we notice that the potential becomes flatter and flatter along ${\cal C}$, and in the whole cavity as well. This is expected because, as $e \rightarrow 0$, curvature effects become less and less pronounced. The shell becomes, locally, similar to a hollow filament, for which the interior potential is a constant by virtue of the Gauss theorem.

\subsection{Shells with a thick edge, i.e. tubes}
\label{subsec:tubes}

If we concatenate coaxial toroidal shells having the same main radius $\rc$ but a different core radius $b \in [\bin,\bout]$, then one gets a toroidal shell with thick edge, i.e. a tube. This system is depicted in Fig. \ref{fig:scheme3.eps}. Each shell has its own pole. It follows that there is no pole for the tube (otherwise, this would imply that $b/\rc$ is constant for each constitutive shell). The total potential is obtained from Eq. (\ref{eq:psihtorus}) by integration over $b$, i.e. 
\begin{flalign}
 \Psi(R,Z) = -2G \int_\bin^\bout{ \rho (b) bdb \int_0^{2\pi} { \sqrt{\frac{a}{R}}  k \elik(k) d\theta }},
\label{eq:psi_thickedge}
\end{flalign}
where $\rho$ is the local mass density. Figure \ref{fig:pot_thick.ps} shows the potential (in units of $GM/\rc$) computed by direct integration\footnote{We have used the trapezoidal rule as the quadrature scheme both in $\theta$ as before, and in $b$ with $N_b=257$ equally spaced nodes.} for a homogeneous tube with parameters $\bin/\rc=0.05$ and $\bout/\rc=0.1$. Figure \ref{fig:pot_all.ps} is for the equatorial plane only. The mass of the tube is
\begin{equation}
    M= 2 \pi^2 \rho (\bout^2-\bin^2) \rc.
    \label{eq:mass_tube}
\end{equation}
We observe the rounded shape of the potential inside the tube. There is no more jump in the derivative of the potential when crossing the system boundaries, as expected for any three-dimensional distribution. At short radii, the potential decreases with $R$. This continues when passing through the outer shell at $R=\rc-\bout$. The minimum (i.e. the place where the acceleration is zero) is reached inside matter, just before crossing the inner shell at $R=\rc-\bin$. Next, the potential increases with $R$ in the cavity, and again inside matter, and this is so up to infinity. Inside the internal cavity, we notice the quasi-linear behavior of $\Psi$ with $R$, which is a consequence of what has been observed for the shell. This is basically the expression of the superposition principle. The central cavity is actually common to all shells the tube is made of. This means that all the approximations presented in Sects. \ref{sec:intpot} and \ref{sec:higherorders} are valid in this common cavity, i.e. for $R \in [\rc-\bin,\rc+\bin]$

\begin{figure}
\includegraphics[width=8.5cm,bb=-50 0 398 277,clip=true]{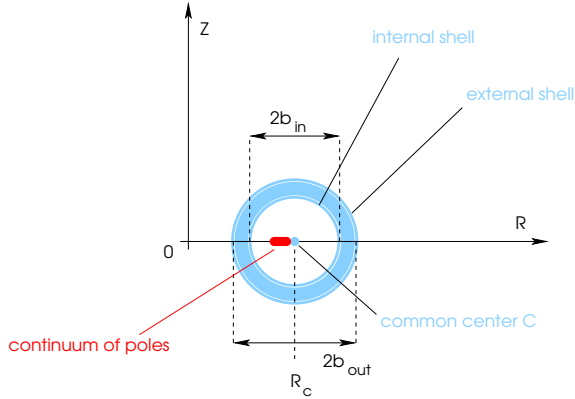}
\caption{The shell with a thick edge (i.e. the tube) is defined by an internal toroidal shell with radius $\bin$ and an external shell with radius $\bout$. Each shell has its own pole, resulting in a continuum of poles ({\it red}).}
\label{fig:scheme3.eps}
\end{figure}

\begin{figure}
\includegraphics[width=6.8cm,bb=50 55 554 770,clip=true,angle=-90]{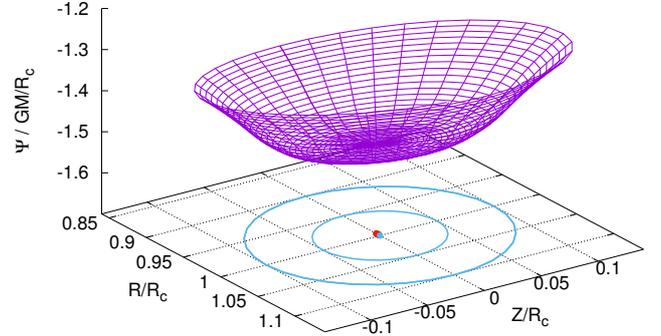}
\caption{The legend is the same as for Fig. \ref{fig:pot.ps} but here the shell has a thick edge with $\bin/\rc=0.05$ and $\bout/\rc=0.1$; see also Fig. \ref{fig:scheme3.eps}. The two limiting sections are drawn ({\it blue lines}) as well as the common center ({\it blue dot}) and the continuum of poles ({\it red dots}).}
\label{fig:pot_thick.ps}
\end{figure}

\begin{figure}
  \includegraphics[height=8.7cm,bb=50 50 554 770,clip=true,angle=-90]{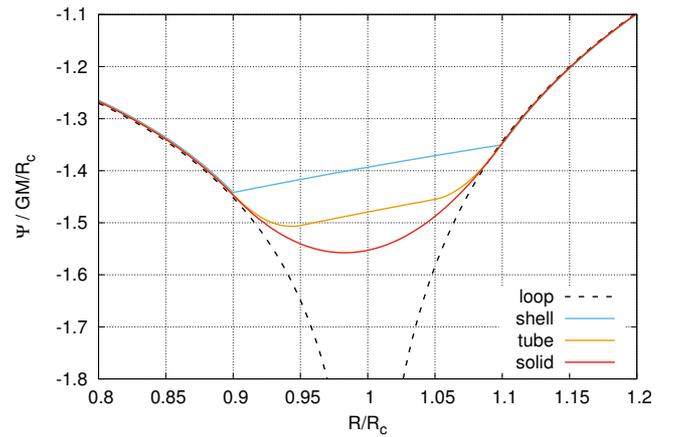}
\caption{Gravitational potential at the equatorial plane for the loop, the shell ($e=0.1$), the tube ($\bin/\rc=0.05$ and $\bout/\rc=0.1$) and the solid torus ($\bout/\rc=0.1$) in units of $GM/\rc$. The axis is in units of $\rc$.}
\label{fig:pot_all.ps}
\end{figure}

How does it work? For instance, if we consider order $0$, we have from Eq. (\ref{eq:ax0pot})
\begin{flalign}
    \Psi & \approx \int_\bin^\bout{ C_0(b) \frac{\rho(b)}{\Sigma} db}.
\end{flalign}
It is important to realise that $C_0 = \psif$ is a function of $b$, not only through the prefactor $-G \Sigma b$, but also through $\kc$. We see that $\rho$ can also depend on $b$ if the core is stratified, and there is no difficulty there. By using Eq. (\ref{eq:psihtorus5}) and changing $b = e \rc$ for $p$ from Eq. (\ref{eq:p}), we obtain
\begin{flalign}
  \label{eq:psi_thickedge2}
  \Psi & \approx  -16 G \rho \int_\bin^\bout{ b \elik(p') \elie(p) db}\\
    \nonumber
  & = -64 G \rho \rc^2 \int_\pin^\pout{  \frac{p^3 \elik(p') \elie(p)}{(1+{p'}^2)^3}  dp}.
\end{flalign}
where $\pin$ and $\pout$ are found from Eq. (\ref{eq:p}),  and $\rho=$const. is assumed. This integral, which can also be written in terms of $\kc$ or $\kpc$, has been found nowhere in the literature \citep[however, see][]{Prudnikov88,WAN2012121}. It must therefore be estimated numerically. If we consider order $1$ for approximating $\Psi$, then we have from Eq. (\ref{eq:ax1pot})
\begin{flalign}
  \Psi & \approx \int_\bin^\bout{ C_0(b)\frac{\rho(b)}{\Sigma}db} + \ln R \int_\bin^\bout{ C_1(b) \frac{\rho(b)}{\Sigma}db}.
\end{flalign}
There is no special difficulty to proceeding to next orders, which has been checked. The results for the tube fully agree with what is reported in Sect. \ref{sec:effectofaxisratio} for the shell. Note that if we set $\bin \rightarrow 0$, then the cavity gets smaller and smaller. The domain of validity of the above formula gets shorter and shorter. For $\bin=0$, the cavity even disappears and the tube becomes a solid torus. Only the potential at $\rc$ becomes accessible.

\subsection{An empirical law for the exterior potential}
\label{subsec:extpot}

The potential outside the cavity of the toroidal shell is not accessible by the formula derived above but it is clearly needed. This would ensure a full coverage of space, which is necessary for dynamical studies \citep{skh04,bvs12}. We show in Fig. \ref{fig:pot_all.ps} the equatorial potential in units of $GM/\rc$ for the four systems considered in this article, i.e. the loop, the shell, the tube and the solid torus. We used the same radius $\rc$ and the same core radius $\bout/\rc=0.1$. Since the mass densities $\lambda$, $\Sigma$ and $\rho$ have been set to unity, these systems have not the same mass. We notice that, while interior potentials are very different, exterior potentials superimpose with each other remarkably. This interesting fact has been outlined in \cite{ba11}. However, the graph above can be misleading. The concordance of exterior potentials is true for small axis-ratios only and fails for $e \gtrsim 0.3$ typically. So, if the condition $e^2 \ll 1$ is satisfyied, the potential outside the cavity of the toroidal shell is close to Eq. (\ref{eq:psiloop}) when appropriately scaled. When coupled to one of the approximations given in the previous sections, one gets a full coverage of the physical space. In summary, by using the $1$st-order approximation for the interor solution (higher-orders can be considered instead), we have
\begin{flalign}
  \frac{\Psi(R,Z)}{ -\frac{GM}{\rc}} \approx
    \label{eq:psi_full}
  \begin{cases}
    \frac{1}{\pi} \sqrt{\frac{\rc}{R}} k_0 \elik(k_0), \\
    \qquad \text{with } k_0^2=\frac{4R \rc}{{(R+\rc)^2+Z^2}}\\
    \qquad \eta < \etac \quad \text{(outside the cavity)},\\\\
   \left[\acal_{00}(\kc) + \acal_{10}(\kc)  \ln \frac{R}{\rf} \right] \cdot \frac{\bcal(\kc)}{4 \pi^2}\\
      \qquad \eta > \etac \quad \text{(inside the cavity)},
    \end{cases}
\end{flalign}
where $M$ is given by Eq. (\ref{eq:mass}) and $k_0$ comes from Eq. (\ref{eq:k}) with $a=\rc$ and $z=0$. A simple program is given in the Appendix \ref{sec:f90program2} for Eq. (\ref{eq:psi_full}).

Besides, the knowledge of the exterior potential would enable us to determine the potential of the solid torus, which is composed of the interior potential of a tube and the exterior potential of a smaller solid torus, {\bf i.e. $\int_0^\bout \dots db' = \int_0^b \dots db' + \int_b^\bout \dots db'$}. From this point of view, an interesting and probably more sraightforward option would be to use the contour integral as the starting point \citep{ansorg03,thh2014Cemda}. This question is open.

\section{Discussion}
\label{sec:discussion}

\subsection{About the implementation}

In general, an analytical approach is more powerful than a fully numerical treatment, mainly because one captures the sensitivity of all the parameters involved on the resulting quantity. One has, however, to make sure that the implementation is not too tricky, and that the advantages in terms of precision and computing time are real. This is especially true when dealing with special functions and series as it is the case here. For instance, expanded Green functions are very often used in potential theory but their efficiency is not always satisfactory \citep{clement74,thh2014Cemda}

We first notice that the differences between the four approximations are indeed minor. For a given value of the shell axis-ratio $e$, only $4$ evaluations of complete elliptic integrals and a few rational functions of $k^2$ are needed to form $\vec{u}$ and $\vec{f}_{ij}$ respectively. These quantities, and subsequently the four constants $C_0$, $C_1$, $C_2$ and $C_4$, can be computed once for all. The potential at a single point $(R,Z)$ of the cavity is then obtained by computing only a few polynomials in $R$ and $Z$ in addition. The comparison with the direct numerical integration is then straightforward. By using the trapezoidal rule with $N_\theta+1$ equally spaced nodes, the number of evaluations of $\elik(k)$ amounts to $N_\theta$ (values at $\theta=0$ and $\theta=2\pi$ are identical). This quadrature scheme being second-order accurate in the grid spacing $2\pi/N_\theta$, the absolute error is $E \sim 2\pi^3/3 N_\theta^2$. So, $4$ different evaluations of the $\elik(k)$ means $N_\theta=4$. The corresponding error is therefore of the order of unity, i.e. much larger than one percent which is the typical value obtained at order $0$ (see Fig. \ref{fig:errvsaprim.ps} for $e \approx 0.01$). In terms of equivalent error, we need $N_\theta \sim \sqrt{2\pi^3/3 E}$ to reach a given error level $E$. This gives $N_\theta \sim 45$ for $E=0.01$ and $N_\theta \sim 450$ for $E=0.0001$ (for $e \approx 0.001$). Since the computing time is mainly governed by the determination of the elliptic integrals, we see that the direct numerical integration is, {\it by orders of magnitudes}, always inferior to the analytical approach, which is what is observed in practice (see Appendix \ref{sec:f90program} for a possible implementation).

\subsection{From shells to fluid tori}

The results presented in this paper belong to the domain of classical theoretical physics. The derivation of the (vacuum) gravitational potential inside the geometrically thin massive shell, as discussed in the present paper, is a first step towards a complete solution for the potential within the matter distribution of the fluid torus. This is a longstanding and recurrent challenge in Astrophysics \citep{dyson1893b,dyson1893a,vl09,fukushima10,ba11}, and in Electrostatics as well \citep{bb83,ha04,sw05,ma18}. The knowledge of the gravity field of toroidal systems is important to derive the equilibrium structure, the shape of the surface, and the their time evolution. It is also fundamental to investigate the dynamics of particles travelling around \citep[e.g.][]{skh04,subrkaras05}. From this point of view, Eq. (\ref{eq:psi_full}a), although empirical, enables such a study.

Any analytical solution readily available can be useful as a test bed for numerical approaches to more astrophysically realistic systems. Actually, we can envisage two types of objects where massive toroidal configurations are relevant. Gravitating very massive tori have been considered as a transient stage during the merger process and the associated tidal disruption event of a neutron star in the close binary system. The remnants can spread and form a toroidal structure around the tidal radius, which eventually becomes partly accreted onto the central body (presumably a black hole) and partly expelled in the form of jet or a massive outflow \citep[e.g.][]{akl98,mne98,lee01}. Besides, gamma rays and neutrinos have been proposed to originate from the short-lived dense torus \citep{woo93,ja14}. At much lower scales (a factor $10^3$ to $10^6$ typically), the formation of the Moon has probably involved a transient ring-like structure of gas, dust and peebles after the early Earth has been impacted by a large body \cite{ls17}. On length-scales larger by a factor $10^6$ to $10^9$, massive tori are believed to orbit supermassive black holes in active galactic nuclei \citep[e.g.][]{goo03,khs04}. Even though the total mass of the torus is thought to be less than the central mass concentrated in SMBH in most observed systems, it has been clearly demonstrated that self-gravity must plays a significant role \citep{coza08,hure00}.

\subsection{Concluding remarks and perspectives}

 The article present a novel contribution to the precise construction of the gravitational potential in toroidal systems. Seen from different angles, both the analytical approaches as well as the numerical computations are rather heavy and the evaluations are always cumbersome. It is thus very useful to develop alternative approaches and simplified models based either on course grids in the numerics or fitting formulas in the analytics; here we attempt to contribute to the latter. We have shown that the potential and its successive derivatives happen to be analytical at the pole of toroidal coordinates for a toroidal shell with circular radius and circular section. On these grounds, we have proposed a new kind of expansion for the potential in the cavity of a shell. We have determined the leading terms up to order $3$. The corresponding approximations are of great precision when compared to the direct numerical integration. This is for instance $8$-digit at order $3$ for a shell with an axis ratio of $e=0.1$. Another major result concerns Newton's theorem for the ellipdoidal homoeoid which cannot be transposed to the toroidal case. The existence of the curvature (around the $z$-axis) makes the potential well deeper at the inner edge of the shell than at the outer edge. Matter is more concentrated at short radii.

This work adresses new questions and requires further developments. While the level of accuracy reached is already very high with te $3$rd-order formula, there is a priori no limit in including more terms in the expansion of the interior potential, but an increasing effort is required to derive new partial derivatives evaluated at the pole. It would be interesting to understand the spike pattern that the error maps exhibit. This is probably the emergence of modes in $\xi$ related to the solution of the Laplace equation in toroidal coordinates with variable seperation \citep[see ][]{ma18}. This point remains to be clarified. Finally, this approach can be transposed to the case of a toroidal current density relevant in astrophysical and laboratory plasmas \citep{slany13}. Actually, in the case of axial symmetry, the vector potential $\mathbfit{A}$ for the current loop is quite similar to Eq. (\ref{eq:psiloop}) \citep[see][]{jackson98,ct99}. This would give access to the poloidal component of the magnetic field.

\section*{Acknowledgments}

J.M. Hur\'e is grateful to the Astronomical Institute in Prague where this work has been initiated during winter 2016 and fundings for the visit. A. Trova acknowledges support from the Research Training Group 1620 ``Models of Gravity'' funded by the German Science Foundation DFG. V. Karas thanks the Czech. Science Fundation (ref. 19-011373). We thank V. Bruneau, N. Popoff at the IMB and B. Boutin-Basillais for fruitful discussions. The anonymous referee is acknowledged for the constructive feedback.

\bibliographystyle{mn2e}

\appendix

\section{Modulus transformations}
From \cite{gradryz07}, we have
\begin{equation}
   \frac{(1+k')}{2}\elik(k)=\elik\left(\frac{1-k'}{1+k'} \right),
  \label{eq:kk2kprim}
\end{equation}
\begin{equation}
  \elie\left(\frac{2\sqrt{k'}}{1+k'}\right)=\frac{1}{1+k'}\left[2\elie(k')-k^2\elik(k') \right],
  \label{eq:ep2kprim}
\end{equation}
and
\begin{equation}
 \elik\left(\frac{2\sqrt{k'}}{1+k'}\right) =(1+k')\elik(k').
  \label{eq:kp2kkprim}
\end{equation}

\section{Derivatives of $\elik$ and $\elie$}

From \cite{gradryz07}, we have
\begin{equation}
    \frac{d\elie(k)}{dk} = \frac{\elie(k)-\elik(k)}{k},
    \label{eq:deoveerdk}
\end{equation}
and
\begin{equation}
    \frac{d\elik(k)}{dk} = \frac{1}{k{k'}^2}\left[\elie(k)-{k'}^2\elik(k)\right].
    \label{eq:dkoveerdk}
\end{equation}

\newpage
\onecolumn
\section{F90 program for the interior potential}
\label{sec:f90program}
\begin{verbatim}
Program F90drivercode
  ! "Interior potential of a toroidal shell from pole values"
  ! Hure, Trova, Karas & Lesca (2019), MNRAS
  ! gfortran F90drivercode.f90; ./a.out
  Implicit None
  Integer,Parameter::SP=Kind(1.00E+00),DP=Kind(1.00D+00),QP=Kind(1.00Q+00)
  Integer,Parameter::AP=DP
  Real(Kind=AP),Parameter::PI=ATAN(1._AP)*4
  Real(KIND=AP),Parameter::EPSMACH=Epsilon(0._AP)
  Real(KIND=AP)::B,RC,MASS,RP ! core radius, main radius and mass of the shell, and position of the pole
  Real(KIND=AP)::KMODC,KMODC2,KPRIMC,KPRIMC2 ! various modulus
  Real(KIND=AP)::R,Z,PSI ! cylindrical coordinates and potential value where it is estimated
  Real(KIND=AP)::PSIP,DPSIDRP,D2PSIDZ2P,D3PSIDZ2DRP ! partial derivatives of the potential at the pole
  Real(KIND=AP)::C0,C1,C2,C4 ! coefficients of the expansion, Eq.(56)
  Real(KIND=AP),Dimension(1:4)::UVECT ! components of the u-vector, Eq.(24)
  Real(KIND=AP),Dimension(1:4)::FVECT00,FVECT10,FVECT02,FVECT12
  Real(KIND=AP)::ELLIPTICK,ELLIPTICKPRIM,ELLIPTICE,ELLIPTICEPRIM
  ! ? input parameters (properties of the shell)
  B=0.1_AP
  RC=1._AP
  MASS=B*RC*PI**2*4
  print*,"Mass of the shell",MASS
  ! statements
  KPRIMC=(Sqrt(RC+B)-Sqrt(RC-B))/(Sqrt(RC+B)+Sqrt(RC-B))
  KMODC2=Sqrt((RC-B)*(RC+B))/(Sqrt(RC+B)+Sqrt(RC-B))**2*4
  KPRIMC2=1._AP-KMODC2
  KMODC=Sqrt(KMODC2)
  RP=RC*KMODC2/(2._AP-KMODC2)
  print*,"Radius of the pole",RP
  ! components of the u-vector, Eq.(24) - values of K(k=KMODC), K(k'), E(k) and E(k') to be set !
  UVECT(1:2)=ELLIPTICK*(/ELLIPTICKPRIM,ELLIPTICEPRIM/)
  UVECT(3:4)=ELLIPTICE*(/ELLIPTICKPRIM,ELLIPTICEPRIM/)
  ! components of the f_ij-vectors, Eq.(26), (32), (51) and (65)
  FVECT00(1:4)=(/-KMODC2,2._AP,0._AP,0._AP/)*8
  FVECT10(1:4)=(/KMODC2*KPRIMC2,-KPRIMC2*2,-KMODC2,1._AP+KPRIMC2/)/KPRIMC2*4
  FVECT02(1:4)=(/-KPRIMC2,KPRIMC2,2._AP+KPRIMC2,-(1._AP+KPRIMC2)*2/)*(KMODC2/KPRIMC2)**2
  FVECT12(1:4)=(/KPRIMC2*(-4._AP+KPRIMC2*3-KPRIMC2**2*2),&
       &KPRIMC2*(4._AP-KPRIMC2+KPRIMC2**2*4),8._AP-KPRIMC2*7-KPRIMC2**2*2+KPRIMC2**3*4,&
       &(1._AP+KPRIMC2)*(-8._AP+KPRIMC2*11-KPRIMC2**2*8)/)*KMODC2**2/KPRIMC2**3*2/3
  ! ? values of R and Z where the potential is requested (must be inside the cavity!)
  R=RP
  Z=0.
  If ((R-RC)**2+Z**2-B**2<0._AP) Then
     PSIP=-DOT_PRODUCT(FVECT00,UVECT)*B
     DPSIDRP=-DOT_PRODUCT(FVECT10,UVECT)*B/RP
     D2PSIDZ2P=-DOT_PRODUCT(FVECT02,UVECT)*B/RP**2
     D3PSIDZ2DRP=-DOT_PRODUCT(FVECT12,UVECT)*B/RP**3
     ! order 3, Eq.(56); set C4=0 for order 2, and set C2=0 for order 1, and set C1=0 for order 0 
     C4=-D3PSIDZ2DRP/RP/32
     C2=(-D2PSIDZ2P-C4*RP**2*16)/4
     C1=(DPSIDRP-C2*RP*2-C4*RP**3*4)*RP
     C0=PSIP-C1*LOG(RP)-C2*RP**2-C4*RP**4
     PSI=C0+C1*LOG(R)+C2*(R**2-Z**2*2)+C4*(R**4-R**2*Z**2*8+Z**4*8/3)
     Print *,"Potential value (3rd-order)",PSI,PSI/MASS*RC
  Endif
\end{verbatim}

\newpage
\onecolumn

\newpage
\section{F90 program for the interior and exterior (empirical) potential}
\label{sec:f90program2}
\begin{verbatim}
Program F90drivercode2
  ! "Interior potential of a toroidal shell from pole values"
  ! Hure, Trova, Karas, & Lesca (2019), MNRAS
  ! gfortran F90drivercode2.f90; ./a.out
  Implicit None
  Integer,Parameter::SP=Kind(1.00E+00),DP=Kind(1.00D+00),QP=Kind(1.00Q+00)
  Integer,Parameter::AP=DP
  Real(Kind=AP),Parameter::PI=ATAN(1._AP)*4
  Real(KIND=AP),Parameter::EPSMACH=Epsilon(0._AP)
  Real(KIND=AP)::B,RC,MASS,RP ! core radius, main radius and mass of the shell, and position of the pole
  Real(KIND=AP)::KMODC,KMODC2,KPRIMC,KPRIMC2,K0 ! various modulus
  Real(KIND=AP)::R,Z,PSI ! cylindrical coordinates and potential value where it is estimated
  Real(KIND=AP)::PSIP,DPSIDRP,D2PSIDZ2P,D3PSIDZ2DRP ! partial derivatives of the potential at the pole
  Real(KIND=AP),Dimension(1:4)::UVECT ! components of the u-vector, Eq.(24)
  Real(KIND=AP),Dimension(1:4)::FVECT00,FVECT10
  Real(KIND=AP)::ELLIPTICK,ELLIPTICKPRIM,ELLIPTICE,ELLIPTICEPRIM
  INTEGER::I,J
  ! ? input parameters (properties of the shell)
  B=0.1_AP
  RC=1._AP
  MASS=B*RC*PI**2*4
  print*,"Mass of the shell",MASS
  ! statments
  KPRIMC=(Sqrt(RC+B)-Sqrt(RC-B))/(Sqrt(RC+B)+Sqrt(RC-B))
  KMODC2=Sqrt((RC-B)*(RC+B))/(Sqrt(RC+B)+Sqrt(RC-B))**2*4
  KPRIMC2=1._AP-KMODC2
  KMODC=Sqrt(KMODC2)
  RP=RC*KMODC2/(2._AP-KMODC2)
  print*,"Radius of the pole",RP
  ! components of the u-vector, Eq.(24) - values of K(k=KMODC), K(k'), E(k) and E(k') to be set !
  UVECT(1:2)=ELLIPTICK*(/ELLIPTICKPRIM,ELLIPTICEPRIM/)
  UVECT(3:4)=ELLIPTICE*(/ELLIPTICKPRIM,ELLIPTICEPRIM/)
  ! components of the f_ij-vectors, Eq.(26), (32), (51) and (65)
  FVECT00(1:4)=(/-KMODC2,2._AP,0._AP,0._AP/)*8
  FVECT10(1:4)=(/KMODC2*KPRIMC2,-KPRIMC2*2,-KMODC2,1._AP+KPRIMC2/)/KPRIMC2*4
  ! 51x51 grid on [0,2]x[-1,1], for Eq.(73)
  Do I=0,50
     R=2._AP*I/50
     Do J=0,50
        Z=2._AP*J/50-1._AP
        If ((R-RC)**2+Z**2-B**2<0._AP) Then
           ! Eq.(73b)
           PSI=-MASS/RC*Dot_PRODUCT(FVECT00+LOG(R/RP)*FVECT10,UVECT)/PI**2/4
        Else
           ! Eq.(73a)
           ! value of K(k0) to be set !
           K0=Sqrt(R*RC/((R+RC)**2+Z**2))*2
           PSI=-MASS/Sqrt((R+RC)**2+Z**2)*ELLIPTICK/PI*2
        Endif
        Print*,R,Z,PSI
     Enddo
   Enddo
\end{verbatim}

\end{document}